\documentclass[twocolumn, pra , nofootinbib , superscriptaddress, longbibliography
]{revtex4-1}
\usepackage{graphicx}
\usepackage{amsfonts}
\usepackage{amssymb}
\usepackage{amsthm}
\usepackage{array}
\usepackage{amsmath}
\usepackage{verbatim} 
\usepackage{hyperref}
\usepackage{color}
\usepackage{bbold}
\usepackage{epstopdf}
\usepackage{mathtools}
\usepackage{enumerate}
\usepackage{subcaption}
\usepackage{physics}
\usepackage{subcaption}
\usepackage{soul,xcolor}
\usepackage{newtxtext,newtxmath}
\usepackage{hyphenat}
\usepackage[normalem]{ulem}

\newtheorem*{proposition*}{Proposition}

\newtheorem*{theorem*}{Theorem}

\newtheorem*{corollary*}{Corollary}

\newcommand{\del}[0]{\partial}
\setstcolor{red}

\begin{document}
\title{Variational quantum algorithms to estimate rank, quantum entropies, fidelity, and Fisher information via purity minimization}
\author{Kok Chuan Tan}
\email{bbtankc@gmail.com}
\affiliation{ School of Physical and Mathematical Sciences, Nanyang Technological University, Singapore 637371, Republic of Singapore}
\author{Tyler Volkoff}
\email{volkoff@lanl.gov}
\affiliation{Theoretical Division, Los Alamos National Laboratory, Los Alamos, NM, USA.}

\begin{abstract}
Variational quantum algorithms (VQAs) that estimate values of widely used physical quantities such as the rank, quantum entropies, the Bures fidelity and the quantum Fisher information of mixed quantum states are developed. In addition, variations of these VQAs are also adapted to perform other useful functions such as quantum state learning and approximate fractional inverses. The common theme shared by the proposed algorithms is that their cost functions are all based on minimizing the quantum purity of a quantum state. Strategies to mitigate or avoid the problem of exponentially vanishing cost function gradients are also discussed.
\end{abstract}

\maketitle

\section{Introduction}

In recent years, there has been growing interest and rapid developments in a class of early quantum computing devices collectively referred to as noisy intermediate scale quantum (NISQ) devices \cite{Preskill2018}. NISQ devices are essentially quantum computing hardware that lacks full quantum error correction. Due to the lack of error correction, the gate fidelities and total coherence times available on such devices are limited. This means that NISQ devices are limited to applications that can be performed using short depth quantum circuits, beyond which the measurement output is unreliable. While NISQ devices cannot perform universal quantum computing, it is widely expected that they are sufficient to provide a demonstrable computational advantage over classical computers in the near term \cite{Preskill2012,Arute2019}. 

One approach to developing algorithms suitable for NISQ devices is to consider hybrid quantum-classical algorithms\cite{Endo2021}. Such algorithms seek to lower the overall quantum circuit depth requirements by offloading a portion of the computation to a classical computer. Among these, a class of algorithms called variational quantum algorithms (VQAs) are arguably among the most widely used and promising strategies for designing a NISQ algorithm. In VQAs, a cost function $C(\theta)$ which is difficult to compute classically is estimated using NISQ hardware, while a classical optimization strategy is used to find the optimal parameter $\theta^*$ that minimizes the cost function. Such strategies have been used in applications such as finding approximate eigenstates of Hamiltonians\cite{Peruzzo2014,Nakanishi2019}, recompiling quantum circuits \cite{qaqc} and solving linear algebraic problems \cite{reben,bravo2019variational,linlin}. As a class of quantum algorithms, VQAs will likely continue to be relevant even when fully quantum error corrected devices are widely available, as the solution to many classical and quantum problems are naturally phrased in terms of the minimization of some cost function. 

In this paper, we develop VQAs to estimate a variety of physical quantities such as the rank of a quantum state, the R{\'e}nyi \cite{Muller2013} and Tsallis \cite{Tsallis1988} quantum entropies, the quantum fidelity \cite{watrousbook} and the quantum Fisher information \cite{Tan2019}. Variations of these algorithms are also adapted to perform tasks such as quantum state learning \cite{Lee2018, Chen2020}, and approximating fractional powers and inverses of density matrices. The cost functions of these VQAs are based on minimizing the quantum purity of the system, defined as the quantity $\tr(\sigma^2)$ for any normalized density matrix $\sigma$. The quantum purity has previously been considered as a quantum resource \cite{Horodecki2003}, and is known to bound the maximum entanglement and quantum coherence generated by a quantum circuit \cite{Streltsov2018}. The algorithms developed here suggest that the quantum purity is also an operationally useful concept in the sense that it can be a useful primitive for constructing cost functions for a variety of VQAs.  

The barren plateau landscape (BPL) problem \cite{McClean2018} in relation to our proposed VQAs is also discussed. The BPL problem is said to occur when the expected magnitude of the gradient of the cost function $C(\theta)$ vanishes exponentially with system size, which in turn suggests that exponential resources are required to optimize the parameters. We provide numerical evidence that the BPL problem can be avoided with sufficiently good ansatz design for low rank input states, and also discuss possible strategies to mitigate the BPL problem by adding local terms to the cost function or by adopting correlated ansatz parameters.

\section{Preliminary equations: quantum SWAP test}
The VQAs that we introduce in Sections \ref{sec:vqas} and \ref{sec:alns} utilize a single algorithmic primitive to calculate the cost function, viz., computation of the trace of a product of quantum states. Although our VQAs are agnostic to the details of how this computation is carried out, here we briefly describe a method based on the quantum SWAP test \cite{Buhrman2001}, which can be generalized for $k$ systems \cite{PhysRevB.96.195136,subasiyirka}. Given $k$ input states $\rho_i$ where $i=1,
\ldots, k$, the SWAP test evaluates the value of $\tr(\prod_{i=1}^k \rho_i)$.

Let us first consider the pure input state $ \ket{\psi_1} \ldots \ket{\psi_k}$. In this case, the goal is to evaluate $\tr(\prod_{i=1}^k |\psi_i\rangle \langle \psi_i |) = \braket{\psi_k}{\psi_1} \braket{\psi_1}{\psi_2} \ldots \braket{\psi_{k-1}}{\psi_k} $. We define the cyclic permutation operator $P_k$ acting on $k$ systems, which performs the operation
\begin{align}
P_k \ket{\psi_1} \ldots \ket{\psi_k}  = \ket{\psi_k} \ket{\psi_1} \ldots \ket{\psi_{k-1}}. \label{eq::permOperator}
\end{align} One may verify that the expectation value of $P_k$ for the input state $ \ket{\psi_1} \ldots \ket{\psi_k}$ gives \begin{align}
\expval{P_k} =  \braket{\psi_k}{\psi_1} \braket{\psi_1}{\psi_2} \ldots \braket{\psi_{k-1}}{\psi_k},
\end{align} which is the required expression.

In order to evaluate this expectation value, we append an ancilla initialized in the state $\ket{+}$, where $\ket{\pm}\coloneqq (\ket{0}\pm \ket{1})/\sqrt{2}$. We then apply a controlled unitary $U$ which performs $P_k$ when the ancilla is in state $\ket{1}$ and the identity operation $\openone_d$ when it is in the state $\ket{0}$. Finally, we measure the ancilla qubit by projecting it onto the $P_\pm \coloneqq \ketbra{\pm}$ basis. The corresponding probabilities are given by 
\begin{align}
\mathrm{Prob}(\pm) &= \norm{(\openone_d \pm P_k) \ket{\psi_1} \ldots \ket{\psi_k}}^2/4 \\
&=\frac{1\pm \expval{P_k}}{2},
\end{align} such that $\expval{P_k} = \mathrm{Prob}(+) - \mathrm{Prob}(-)$. This provides a resource efficient method to evaluate $\expval{P_k}$.

We now consider mixed quantum states. Suppose that the input state has the form $\rho_1 \otimes \ldots \otimes \rho_k $, where $\rho_i$, $i=1,\ldots, k$ may be mixed states in general. Let $\{p_{i, j_i}, \ket{e_{i, j_i}}\}$ denote the $j$th eigenvalues and eigenstates of $\rho_i$. 
We can then write 
\begin{align}
\tr(\prod_{i=1}^k \rho_i) &= \sum_{j_1 \ldots j_k} p_{1,j_1}  \ldots p_{k,j_k} 
\times \notag \\
&\quad \tr( P_k\ketbra{e_{1,j_1}} \otimes \ldots \otimes \ketbra{e_{k,j_k}} ),
\end{align} which is a linear sum of expectation values of $P_k$ acting on pure states. The SWAP test therefore also applies to mixed states.

\section{Variational algorithms via purity minimization\label{sec:vqas}}

We utilize the quantum purity as a basic building block to construct efficiently computable cost functions for our VQAs. First,  note that the purity is only nontrivial when optimizing over mixed quantum states, since it is always equals to unity over pure states. As such, a basic assumption we will make in the ensuing discussion is that the input state of the algorithm $\rho$ has $\mathrm{rank}(\rho) \geq 2$, which implies that $\rho$ is a mixed quantum state.

The following is the basic cost function that we will consider:
\begin{align}
C(\theta) = \tr \left \{[ \eta(\theta)^{k}\rho \eta(\theta)^{k}/\tr(\eta(\theta)^{k} \rho \eta(\theta)^k)] ^2 \right \}, \label{eq::costFun}
\end{align} where $k$ is any positive integer, $\rho$ is the density matrix of a possibly unnormalized mixed state, and $\eta(\theta)$ is some normalized ansatz state. The goal is to minimize the cost function $C(\theta)$ by varying $\theta$ in order to find the optimal ansatz $\eta(\theta^*)$. If we let $\sigma = \eta(\theta)^{k} \rho \eta(\theta)^{k}/\tr(\eta(\theta)^{k} \rho \eta(\theta)^k)$, we see that $\sigma$ is a normalized density matrix and the optimization reduces to finding $\min_\sigma \tr(\sigma^2)$, i.e. it is minimizing the purity of $\sigma$. The purity is minimized by the unique solution $\sigma^* = \openone_d/d$, where $d\geq 2$ is the rank of the input matrix $\rho$ and $\openone_d$ is the identity of the subspace spanned by the support of $\rho$. As a result, minimizing the cost function is equivalent to finding the optimal state $\eta(\theta^*) = \rho^{-1/(2k)}/\tr(\rho^{-1/(2k)})$, given that the parameterized ansatz $\eta(\theta)$ includes this state. This can be verified by substituting this expression for $\eta(\theta^*)$ into Eq.~(\ref{eq::costFun}) to obtain $C(\theta^*) = \tr[\openone_d / \tr(\openone_d)^2] = 1/d $. 

We need to further verify that the cost function $C(\theta)$ is efficiently computable on a quantum computer. Evaluating $C(\theta)$ requires one to compute the quantities (i) $\tr[\left (\eta(\theta)^k \rho \eta(\theta)^k \right )^2]$ and (ii) $\tr(\eta(\theta)^k\rho \eta(\theta)^k) = \tr(\rho \eta(\theta)^{2k})$. Both (i) and (ii) may be efficiently computed via the  SWAP test using (i) 2 copies of $\rho$ and $4k$ copies of $\eta(\theta)$ and (ii) a single copy of $\rho$ and $2k$ copies of $\eta(\theta)$ respectively.  In general, the controlled cyclic permutation operator $P_k$ from Eq.~(\ref{eq::permOperator}) acting on $k$ subsystems can be decomposed into a series of $(k-1)$ controlled SWAP operations. Assuming that each of the $k$ subsystems is composed of $n = \log d$ qubits, each of the $k$ controlled SWAPs on the level of subsystems can be implemented using $n$ controlled SWAP operations on the level of individual qubits. Combining these two facts, the circuit complexity of the SWAP test is $\order{k \log d}$, or just $\order{n}$ assuming the value of $k$ is fixed. This shows that the cost function $C(\theta)$ can be efficiently sampled in linear time.

We also observe that one may estimate the normalization factor $\tr(\rho^{-1/(2k)})$ by utilizing the optimal state via the expression $\tr[\rho \eta(\theta^{*})^{2k}] = d/\tr(\rho^{-1/(2k)})^{2k}$, which results in the identity
\begin{align}
\tr(\rho^{-1/(2k)})^{2k} &= [\tr(\rho \eta(\theta^{*})^{2k}) /d]^{-1}\nonumber \\
&= [\tr(\rho \eta(\theta^{*})^{2k}) C(\theta^*)]^{-1}. \label{eq::normalization}
\end{align}

Away from the critical point $\theta^{*}$, the product $[\tr(\rho \eta(\theta)^{2k}) C(\theta)]^{-1}$ provides an approximation of the normalization factor.

\subsection{Geometric interpretation of purity minimization}

We highlight that quantum purity has a direct geometric interpretation. Let us consider the Hilbert-Schmidt norm $\lVert A \rVert \coloneqq \sqrt{\tr(A^\dag A)} $ and the corresponding norm induced distance $d(A , B) \coloneqq \lVert (A-B) \rVert$. We compute the square distance between the maximally mixed state and a normalized quantum state $\sigma$:
\begin{align}
d(\frac{\openone_d}{d} , \sigma)^2 &= \tr[(\frac{\openone_d}{d} - \sigma)^2] \\
&= \tr(\frac{\openone_d}{d^2} + \sigma^2 -2 \frac{\sigma}{d}) \\
&= \tr (\sigma^2) - \frac{1}{d}. \label{eq::geomInt}
\end{align}

From this, it is observed that the quantum purity directly quantifies how far away $\sigma$ is from the maximally mixed state, as quantified by the Hilbert-Schmidt distance $d(A,B)$. By defining $\sigma(\theta) \coloneqq  \eta(\theta)^{k} \rho \eta(\theta)^{k}/\tr(\eta(\theta)^{k} \rho \eta(\theta)^k)$, we have $\tr [\sigma(\theta)^2] = C(\theta)$, the cost function in Eq.~(\ref{eq::costFun}). Minimizing $C(\theta)$ therefore has a direct interpretation of finding the geometrically closest state $\sigma(\theta)$ on the ansatz manifold from the the maximally mixed state $\openone_d/d$.

\section{Applications of purity minimization\label{sec:alns}}

It turns out that purity minimization is a useful primitive to solve a range of problems that are relevant in quantum physics. We will demonstrate this by describing several possible applications based on the cost function described in Eq.~(\ref{eq::costFun}). 

\subsection{Rank estimation} One immediate application is to estimate the rank of the input state $\rho$. This can be useful in, for instance, quantum state tomography, where efficient measurement strategies can be devised for low rank states\cite{ODonnell2016}. Purity optimization allows one to find a lower bound estimate of the rank. This comes directly  from the fact that $d = \mathrm{rank}(\rho)\approx 1/C(\theta^*)$ and that in general for any $\theta$, we have $d \geq 1/C(\theta)$.  For rank estimation, the size of the quantum circuit required scales with $5n+2$, assuming that the input states are given and the dimension of $\rho$ is $d= 2^n$. The overall quantum circuit complexity is  $\order{n}$.

\subsection{Approximating fractional powers and fractional inverses of density matrices}
Suppose we would like to raise the power of the density matrix to some power $\rho^\alpha$, for some real value $\alpha \in [-1,1]$. We show that this state can be approximately prepared via purity minimization.

We assume that $\alpha$ is approximated by some rational number $\pm p/q$, where $p,q$ are both positive integers and $p \leq q$. The case where $\alpha = 1$ is trivial since we assumed that $\rho$ is a given input state. The other trivial case is when $\alpha = 0$, where the solution is just the maximally mixed state regardless of the input state $\rho$.

For the case $\alpha = -1$, we note that this is approximated by minimizing the purity of the (unnormalized) state $\eta(\theta_1) \rho^2 \eta(\theta_1)$. This finds a solution $\eta(\theta_1^*)$ such that $\eta(\theta_1^{*}) \rho^2 \eta(\theta_1^{*})$ is proportional to the identity. This implies that $\eta(\theta_1^{*}) = \rho^{-1}/\tr(\rho^{-1})$. Recall that the denominator can be estimated using Eq.~(\ref{eq::normalization}).

Next, we consider the case where $\alpha > 0 $ and $p < q$. By choosing $k=q$ in Eq.~(\ref{eq::costFun}), we find a state  $\eta(\theta_1^*) \approx \rho^{-1/(2q)}/\tr(\rho^{-1/(2q)})$. The next step is to perform another round of purity minimization for the (unnormalized) input state $\mu(\theta_2) \eta(\theta_1^*)^{4p} \mu(\theta_2)$, which finds some optimal state $\mu(\theta_2^*)$. Since for a well-chosen ansatz, $\eta(\theta_1^*)^{4p} $ is approximately $[\rho^{-1/(2q)}]^{4p} = \rho^{-(2p)/q}$ up to a normalization factor, we have $\mu(\theta_2^*) \approx \rho^{p/q}/\tr(\rho^{p/q})$ when $\mu(\theta_2^*) \eta(\theta_1^*)^{4p} \mu(\theta_2^*)$ is proportional to the identity.

We now consider the case $\alpha < 0 $ and $p < q$. For this, we will use the state $\mu(\theta_2^*)$ from the preceding paragraph as the input. We then perform another round of purity minimization for the (unnormalized) state $\nu(\theta_3)\mu(\theta_2^*)^2\nu(\theta_3)$. This finds the optimal state $\nu(\theta_3^*)$ so that $ \nu(\theta_3^*)\mu(\theta_2^*)^2\nu(\theta_3^*) $ is proportional to the identity, which implies $\nu(\theta_3^*) \approx \rho^{-p/q}/ \tr(\rho^{-p/q})$. This covers all the important cases and is sufficient to show that purity optimization is able to approximate the state $\rho^\alpha$, for any $\alpha \in [-1,1]$.

For approximating fractional powers and fractional inverses, the size of the largest quantum circuit required scales with $\max ((8p+3)n+2, (2q+1)n +2)$, assuming that the input states are given and the dimension of $\rho$ is $d= 2^n$. The overall quantum complexity is therefore $\order{n}$ for fixed $p,q$.

\subsection{Quantum state learning}

The goal of quantum state learning\cite{Lee2018, Chen2020} is to learn a quantum circuit that produces an approximation of a given input state $\rho$. Since $\eta(\theta) = \tr_b[U(\theta)\ketbra{0}_{ab}U^\dag(\theta)]$ where $U(\theta)$ represents some parametrized quantum circuit, this is equivalent to finding some optimal $\theta^*$ such that $\eta(\theta^*) \approx \rho$ for any target mixed state $\rho$.

In the preceding section, we described how to find fractional inverses of $\rho$. Consider the special case where $\alpha = -1$, and the optimal solution is $\eta(\theta_1^*) \approx \rho^{-1}/\tr(\rho^{-1})$. We then perform another purity minimization over $\theta_2$ for the state
\begin{align}
\frac{\nu(\theta_2)\eta(\theta_1^*)^2 \nu(\theta_2)}{\tr[\nu(\theta_2)\eta(\theta_1^*)^2 \nu(\theta_2)]}.
\end{align} We see that since $\eta(\theta_1^*)^2$ approximates $\rho^{-2}$ up to a normalization factor, the optimal solution is achieved when $\nu(\theta_2^*) \approx \rho$, which is the required state. The size of the largest quantum circuit required to perform quantum state learning is $7n+2$, assuming that the input states are given and the dimension of $\rho$ is $d= 2^n$. The overall circuit complexity is $\order{n}.$

In Ref.~\cite{Chen2020}, a similar technique using the SWAP test was used to compute a different cost function. This leads to a similar circuit complexity of $\order{n}$. However, we note that the purity minimization approach incurs significant additional overhead because two parameters need to be optimized in order to find $\eta(\theta_1^*)$ and $\nu(\theta_2^*)$ while only one parameter needs to be optimized in the method of Ref.~\cite{Chen2020}. 

\subsection{Estimating R{\'e}nyi and Tsallis entropies}

We recall that the R{\' e}nyi entropies\cite{Buhrman2001} are defined as the quantity
\begin{align}
S^R_\alpha (\rho) = \frac{1}{1-\alpha} \log \tr(\rho^\alpha),
\end{align} while Tsallis entropies\cite{Tsallis1988} are defined as
\begin{align}
S^T_\alpha (\rho) = \frac{1}{1-\alpha}  [\tr(\rho^\alpha)-1],
\end{align} where $\alpha \in (0,1)\cup (1,\infty)$.  The widely used von Neumann entropy $S(\rho)\coloneqq -\tr(\rho \log \rho)$ is retrieved as the limiting case of both quantities when $\alpha \rightarrow 1$, i.e. $\lim_{\alpha \rightarrow 1}S^R_\alpha (\rho) = \lim_{\alpha \rightarrow 1}S^T_\alpha (\rho) = S(\rho)$. Note that both quantities require some estimate of $\tr(\rho^\alpha)$. For integer $\alpha$, non-variational quantum algorithms have been proposed \cite{subasicincio,PhysRevB.96.195136}.

In the preceding section, we have previously discussed how to obtain an approximation of fractional powers of a density matrix, i.e. prepare a quantum approximation $\eta(\theta^*) \approx \rho^{p/q}/\tr(\rho^{p/q})$. Observe that any $\alpha \in (0,1)\cup (1,\infty)$ has a rational approximation $l + p/q$, where $l = \lfloor \alpha \rfloor $, $p$ and $q$ are nonnegative integers. This allows us to write $\tr(\rho^\alpha) \approx \tr(\rho^l \rho^{p/q} ) \approx \tr[\rho^l \eta(\theta^*)] \tr(\rho^{p/q})$, where the first approximation is from the rational approximation to $\alpha$ and the second approximation is due to the form of the ansatz. $\tr[\rho^l \eta(\theta^*)]$ can be found by performing the SWAP test using $l$ copies of $\rho$ and one copy of $\eta(\theta^*)$ as input. The normalization factor $\tr(\rho^{p/q})$ can be obtained using information that was already collected during the purity minimization process (see Eq.~(\ref{eq::normalization})). The size of the largest quantum circuit required for this task is $\max(ln+2,(8p+3)n+2, (2q+1)n +2 )$, assuming that the input states are given and the dimension of $\rho$ is $d= 2^n$. The overall circuit complexity is $\order{n}.$
This enables us to estimate both R{\'e}nyi and Tsallis entropies for fractional $\alpha$. By comparison, the algorithms of Refs.~\cite{subasicincio,PhysRevB.96.195136} only applies to integer $\alpha$, with a similar circuit complexity of $\order{n}$.

\subsection{Quantum fidelity and quantum Fisher information estimation}

The Bures fidelity, also called the quantum fidelity, is frequently used to benchmark the quality of a quantum state preparation\cite{watrousbook}. It is also applied in fundamental studies that use geometric quantifiers to detect quantum phase transitions in many body quantum systems\cite{Carollo20201, Tan2020}. A closely related quantity is the quantum Fisher information (QFI). The QFI is most commonly used as a quantifier of the minimal achievable error in a quantum metrology protocol for a quantum probe\cite{holevobook,Tan2019}, but can also be applied in fundamental studies of quantum nonclassicality\cite{Hyllus2012, Tan2018, Tan2019-2}. We recall the expression for the Bures fidelity:
\begin{align}
F(\rho, \sigma) = \tr(\sqrt{\sqrt{\sigma}\rho\sqrt{\sigma}}).
\end{align} From this expression, we expect that if the square root of a density matrix can be estimated variationally, then the quantum fidelity can also be estimated. We will describe how to do this via purity minimization.

First, we choose $k=1$ to find $\eta(\theta_1^*) \approx \sigma^{-1/2}/\tr(\sigma^{-1/2})$. One then uses Eq.~(\ref{eq::normalization}) to obtain an estimate of $\sqrt{\sigma}$ given by the expression 
\begin{align}
\sqrt{\sigma} \approx \sigma \eta(\theta_1) K_{1},
\end{align} where $K_1 \coloneqq  [\tr(\sigma \eta(\theta_1^*)^2) C_1(\theta^*_1)]^{-1/2}$ with $C_1(\theta_{1}^*)$ the value of the cost function evaluated at the optimal ansatz state $\eta(\theta_1^{*})$.

The next step is to perform another purity minimization to find $\nu(\theta_2^*) \approx (\sqrt{\sigma}\rho\sqrt{\sigma})^{-1/2}/\tr[(\sqrt{\sigma}\rho\sqrt{\sigma})^{-1/2}]$. In order to do this, one minimizes the purity of the (unnormalized) state ansatz
\begin{align}
\nu(\theta_2) \sigma \eta(\theta_1^*)\rho \eta(\theta_1^*)\sigma \nu(\theta_2) \label{eq::fidelityState}
\end{align} by optimizing $\theta_2$. We refer to the cost function Eq.(\ref{eq::costFun}) applied to ansatz $\nu(\theta_{2})$ as $C_{2}(\theta_{2})$. Since $ \sigma \eta(\theta_1^*)\rho \eta(\theta_1^*) \sigma  \propto \sqrt{\sigma}\rho\sqrt{\sigma} $, we must have $\nu(\theta_2^*) \propto (\sqrt{\sigma}\rho\sqrt{\sigma})^{-1/2} $. This gives us $\sigma \eta(\theta_1^*)\rho \eta(\theta_1^*)  \sigma \nu(\theta_2^*) \propto \sqrt{\sqrt{\sigma}\rho\sqrt{\sigma}}$. One may verify the final expression
\begin{align}
K_1 K_2\tr[\sigma \eta(\theta_1^*)\rho  \eta(\theta_1^*) \sigma \nu(\theta_2^*) ] &\approx \tr(\sqrt{\sqrt{\sigma}\rho\sqrt{\sigma}}) \label{eq::fidExp}\\
&=F(\rho, \sigma)
\end{align} where $K_2 \coloneqq  [\tr(\sigma \eta(\theta_1^*)\rho  \eta(\theta_1^*) \sigma \nu(\theta_2^*)^2) C_2(\theta^*_2)]^{-1/2}$, with 
$C_2(\theta_2^*)$ the cost function value corresponding to the state 
$\nu(\theta_2^*)$.
Both $K_1$ and $K_2$ can be efficiently estimated using information that was already collected during the purity minimization process (see Eq.~(\ref{eq::normalization})). This allows us to variationally find the quantum fidelity $F(\rho,\sigma)$ using multiple copies of $\rho$ and $\sigma$ as inputs. The space requirements of the protocol are obtained by counting the number of copies of $\rho$, $\sigma$, and variational ansatze needed to compute each factor of the left hand side of Eq.~(\ref{eq::fidExp}). For example, the purity minimization in $K_{1}$ requires 6 input states plus one ancilla qubit to to measure, while $K_2$ requires 14 input states plus one ancilla qubit in total.  For fidelity estimation, the size of the largest quantum circuit scales with $13n+2$, assuming that the input states are given and the dimensions of $\rho$ and $\sigma$ are $d= 2^n$. The overall circuit complexity is $\order{n}$.

In order to approximate the QFI, one can utilize the above purity based VQA for the fidelity. Suppose that a smooth quantum trajectory $\rho_\theta$ in state space is parameterized by the real parameter $\theta$. Assume that every point along this trajectory is given and can be used as an input to some quantum algorithm. One may then approximate the Fisher information via the identity
\begin{align}
I(\theta) = \lim_{\delta \rightarrow 0 } 8 \frac{1-F(\rho_\theta,\rho_{\theta+\delta})}{\delta^2}.
\end{align} Therefore, the QFI can be estimated by variationally optimizing for the quantum fidelity $F(\rho_\theta,\rho_{\theta+\delta})$ between two nearby states $\rho_\theta$ and $\rho_{\theta+\delta}$.  The overall circuit complexity of the QFI estimation is similar to fidelity estimation, assuming that $\rho_\theta$ is given.

 We note that there are several other recently proposed algorithms for estimating the quantum fidelity. Ref.~\cite{Chen2020} contains an alternative VQA approach based on Uhlmann's theorem, which achieves similar overall complexity $\order{n}$. Similar to our approach, it also requires two separate optimization procedures, but our purity minimization procedure incurs significant overhead due to the fact that estimating the purity of Eq.~(\ref{eq::fidelityState}) requires 14 states as input in total. Ref.~\cite{Cerezo2020} describes a VQA that estimates the fidelity by computing upper and lower bounds in $\mathrm{poly}(n)$ time, compared to the $\order{n}$ scaling using purity minimization. Ref.~\cite{Wang2021} describes a fully quantum algorithm that estimates the quantum fidelity of low rank states using block encoding and quantum phase estimation. For low rank states, this achieves an overall complexity of $\mathrm{poly}(n)$ but is more difficult to implement on NISQ devices compared to other VQA approaches due to its heavy reliance on quantum phase estimation. A VQA for finding a lower bound estimate of the QFI is also described in Refs.~\cite{Beckey2020, Sone2021}. This utilizes the fidelity estimation technique of Ref.~\cite{Cerezo2020} as a subroutine and has a circuit complexity of $\mathrm{poly}(n)$. 

\section{Numerical simulation}

\begin{figure*}[t]
	\captionsetup[subfigure]{justification=centering}
	\centering
		\begin{subfigure}[c]{0.4\textwidth}
		\includegraphics[width=\textwidth]{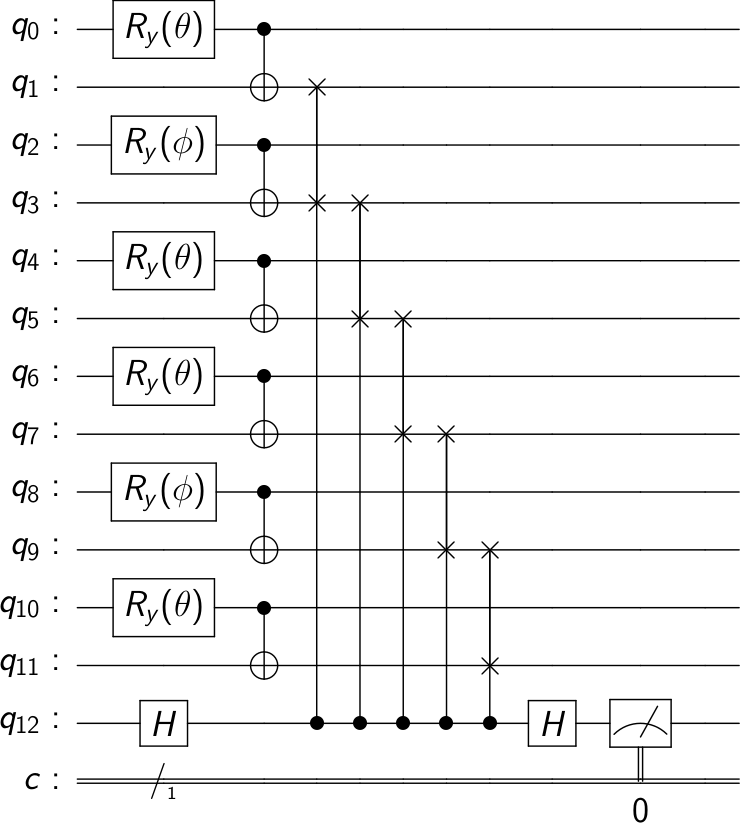}
	\end{subfigure}
		\begin{subfigure}[c]{0.4\textwidth}
		\includegraphics[width=\textwidth]{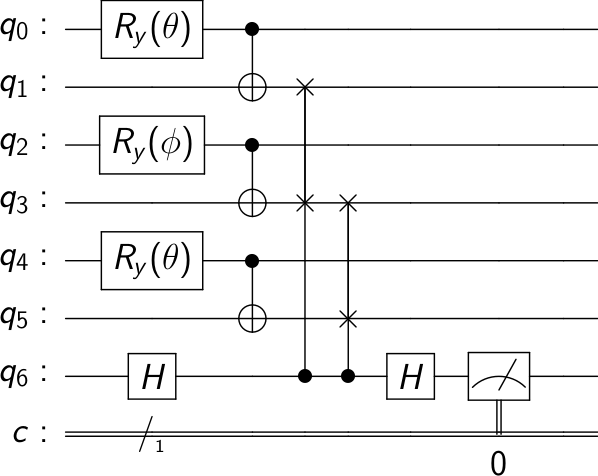}
	\end{subfigure}
\caption{Example circuits to compute (left) the numerator, and (right) the denominator of the cost function $C(\theta)$ with $k=1$ (see Eq.~(\ref{eq::costFun})). In both cases, the output is obtained by measuring a single qubit in the computational basis.}
	\label{fig::circuit}
\end{figure*}

In order to demonstrate the proposed algorithms, we performed a numerical simulation to find the rank, R{\'e}nyi entropy and quantum fidelity for a single qubit input state $\rho = \cos^2(\phi/2) \ketbra{0} +\sin^2(\phi/2) \ketbra{1}$. For this purpose, we chose a simple diagonal ansatz of the form $\eta(\theta) = \cos^2(\theta/2) \ketbra{0} +\sin^2(\theta/2) \ketbra{1}$ initialized at $\theta = \pi/2$ and minimized the cost function over $\theta$ using a gradient descent method.  The gradient of the cost function is computed via a finite difference method using a step size set to $\Delta \theta = 0.005$, while the learning rate of the gradient descent, which determines the rate at which $\theta$ is updated during gradient descent, is set to $r = 0.05$.  The algorithm  is then executed via  simulated quantum circuits using the \textsc{qiskit} quantum simulator package \cite{Qiskit}. Figure~\ref{fig::circuit} shows example circuits for calculating the numerator and denominator of the cost function in Eq.~(\ref{eq::costFun}) with $k=1$. 

\begin{figure*}[t]
	\captionsetup[subfigure]{justification=centering}
    \centering
	\begin{subfigure}[c]{0.32\textwidth}
		\includegraphics[width=\textwidth]{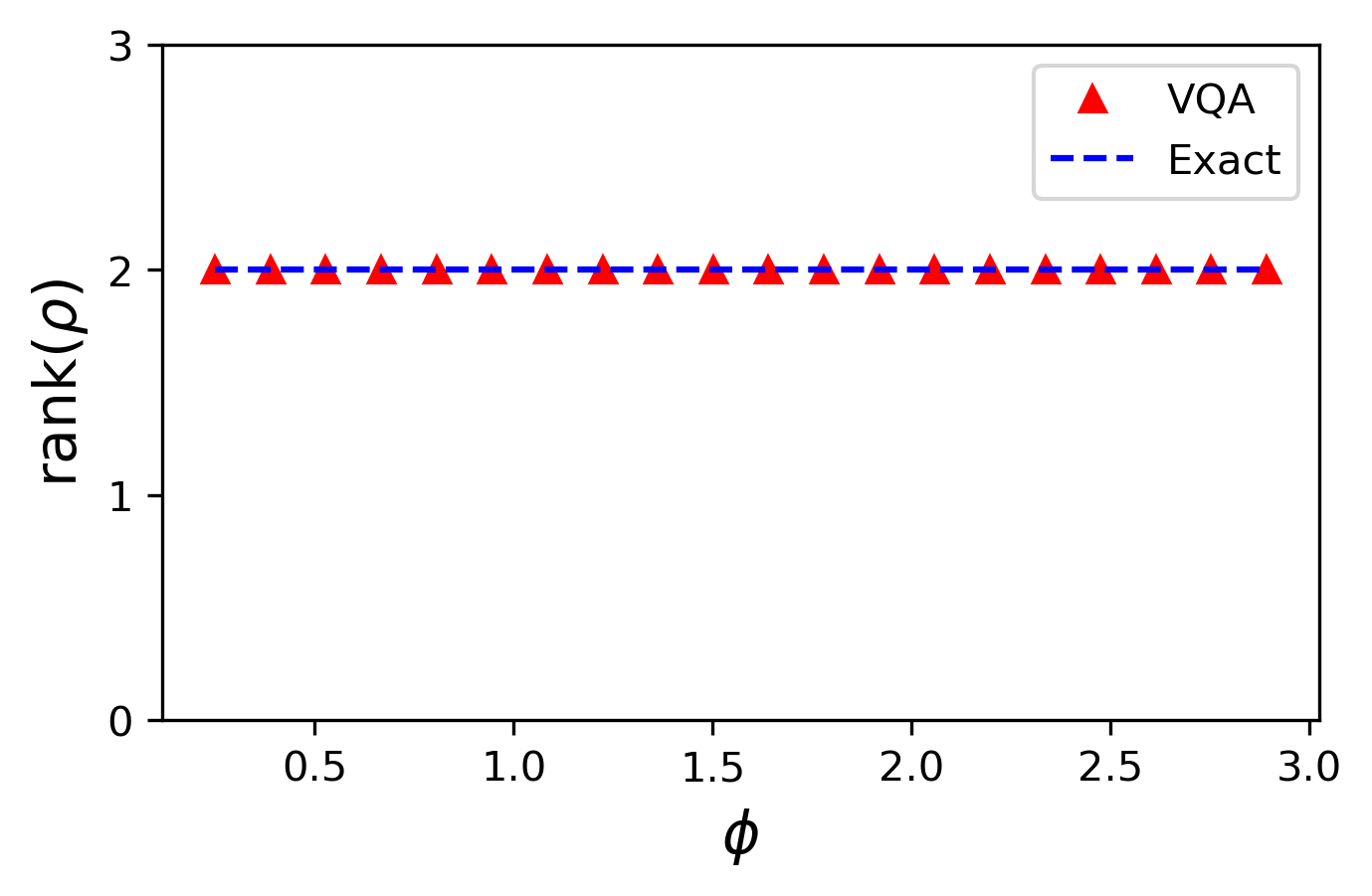}
		
	\end{subfigure}
	\begin{subfigure}[c]{0.33\textwidth}
		\includegraphics[width=\textwidth]{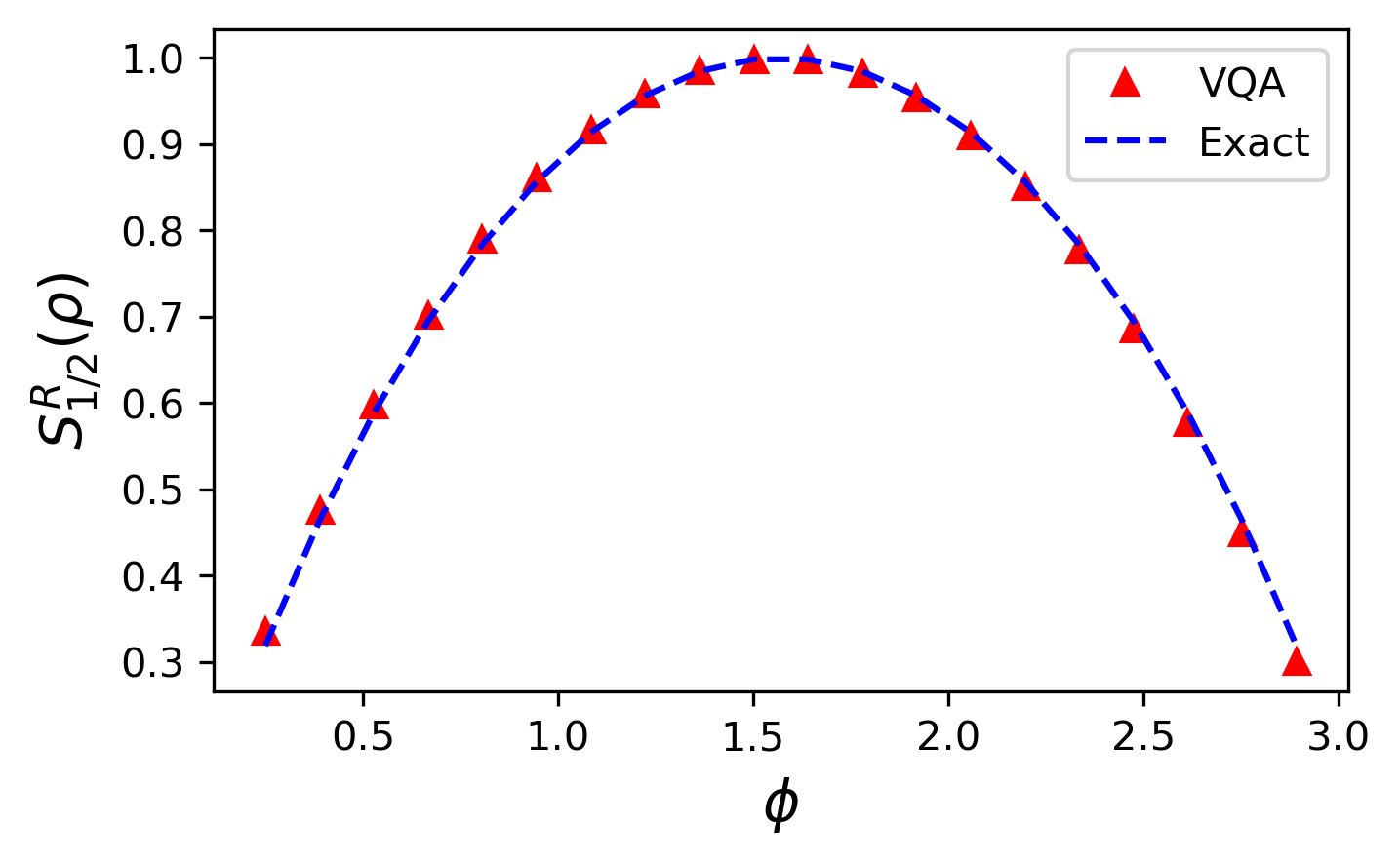}
	
	\end{subfigure}
	\begin{subfigure}[c]{0.33\textwidth}
	    \includegraphics[width=\textwidth]{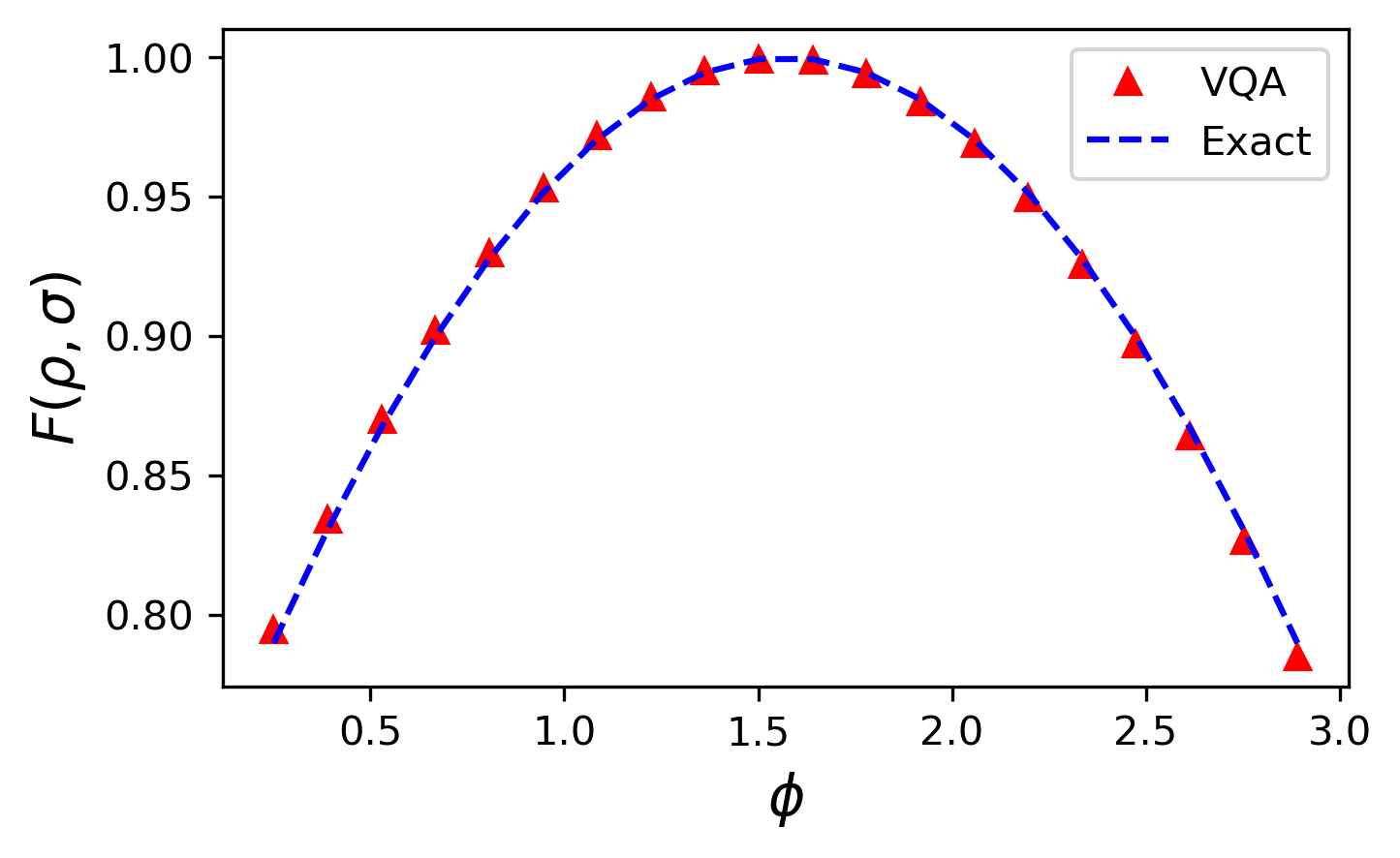}
	    
	\end{subfigure}
	\caption{Exact and estimated values of the (left) rank, (middle) R{\'e}nyi entropy for $\alpha = 1/2$, and (right) fidelity using the proposed VQAs. The input state is parametrized by $\phi$ and is given by $\rho = \cos^2(\phi/2) \ketbra{0} +\sin^2(\phi/2) \ketbra{1}$. For the fidelity calculation $F(\rho,\sigma)$, the second input state was chosen to be the maximally mixed state $\sigma = \openone_2/2$.}
	\label{fig::rankEntropy}
\end{figure*}

We used the proposed algorithms to estimate the rank of $\rho$ as well as R{\'e}nyi entropy for $\alpha = 1/2$. We also used the algorithm to estimate the quantum fidelity $F(\rho, \sigma)$, where the second argument was chosen to be the maximally mixed qubit state, i.e. $\sigma = \openone_2/2$. The results are shown in Figure~\ref{fig::rankEntropy}. It is observed that the algorithms were able to accurately reproduce the expected exact results up to the numerical precision of the optimization. 

\begin{figure*}[t]
	\captionsetup[subfigure]{justification=centering}
    \centering
	\begin{subfigure}[c]{0.32\textwidth}
		\includegraphics[width=\textwidth]{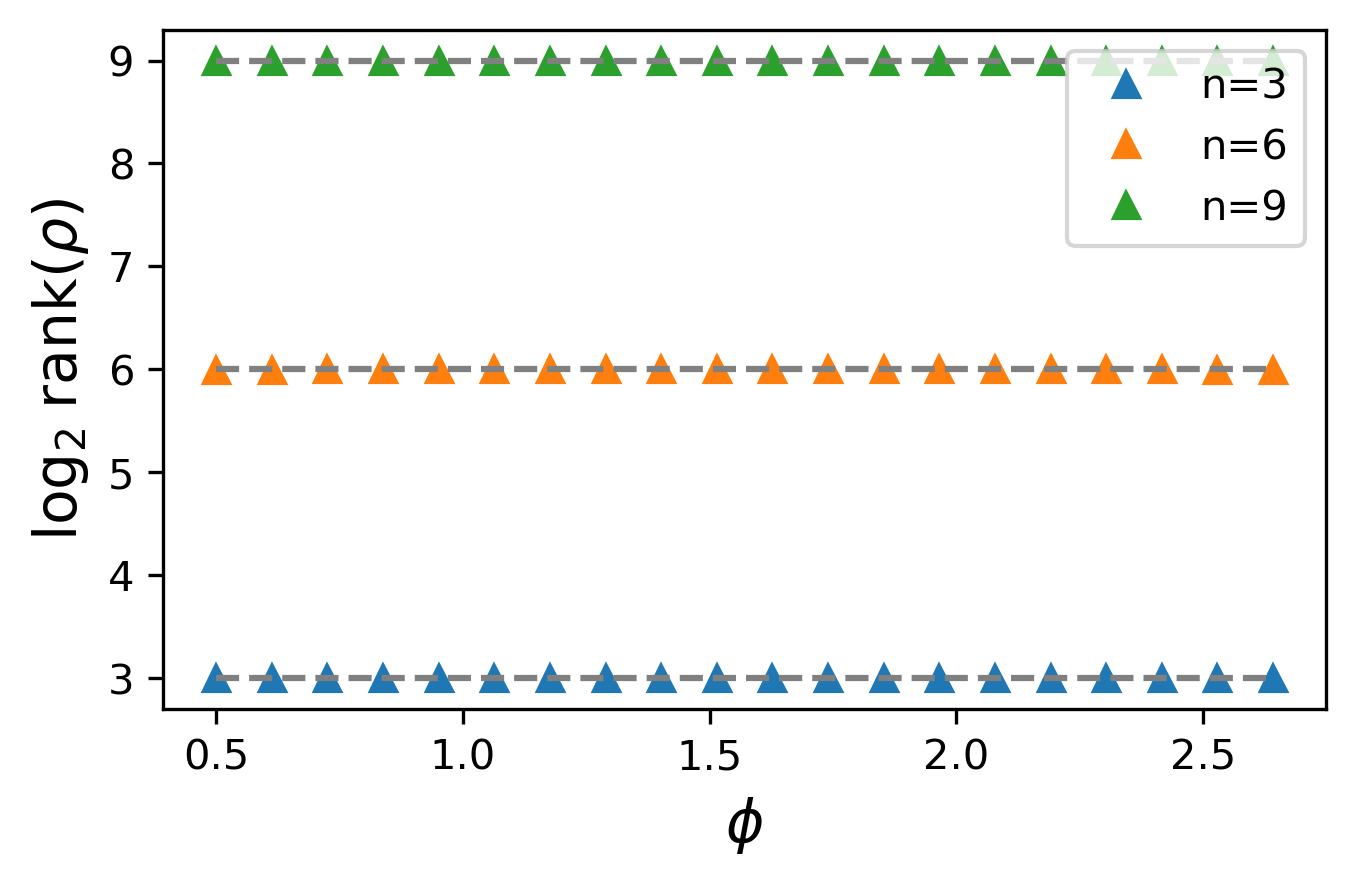}
	\end{subfigure}
	\begin{subfigure}[c]{0.33\textwidth}
		\includegraphics[width=\textwidth]{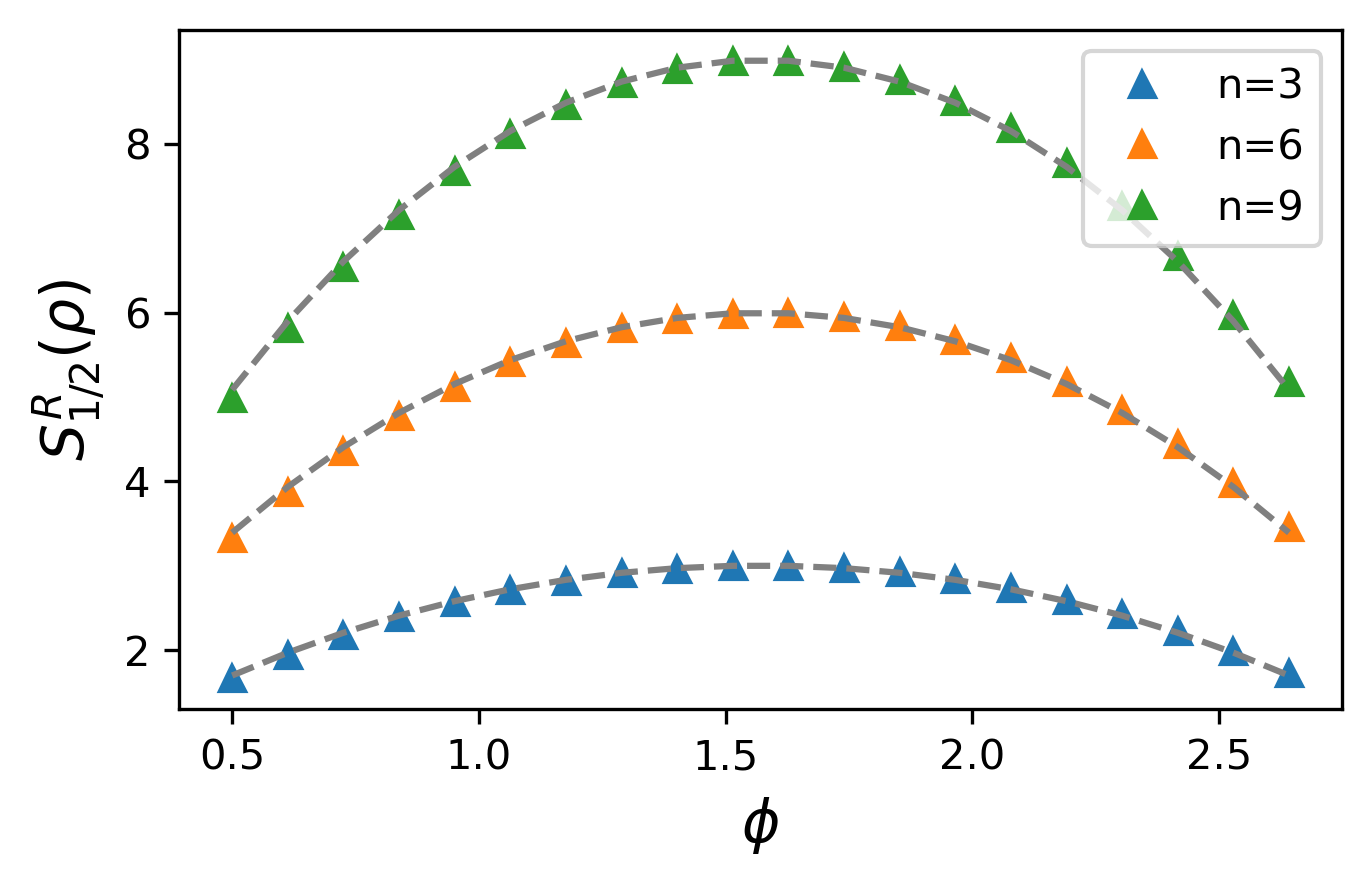}
	\end{subfigure}
	\begin{subfigure}[c]{0.33\textwidth}
	    \includegraphics[width=\textwidth]{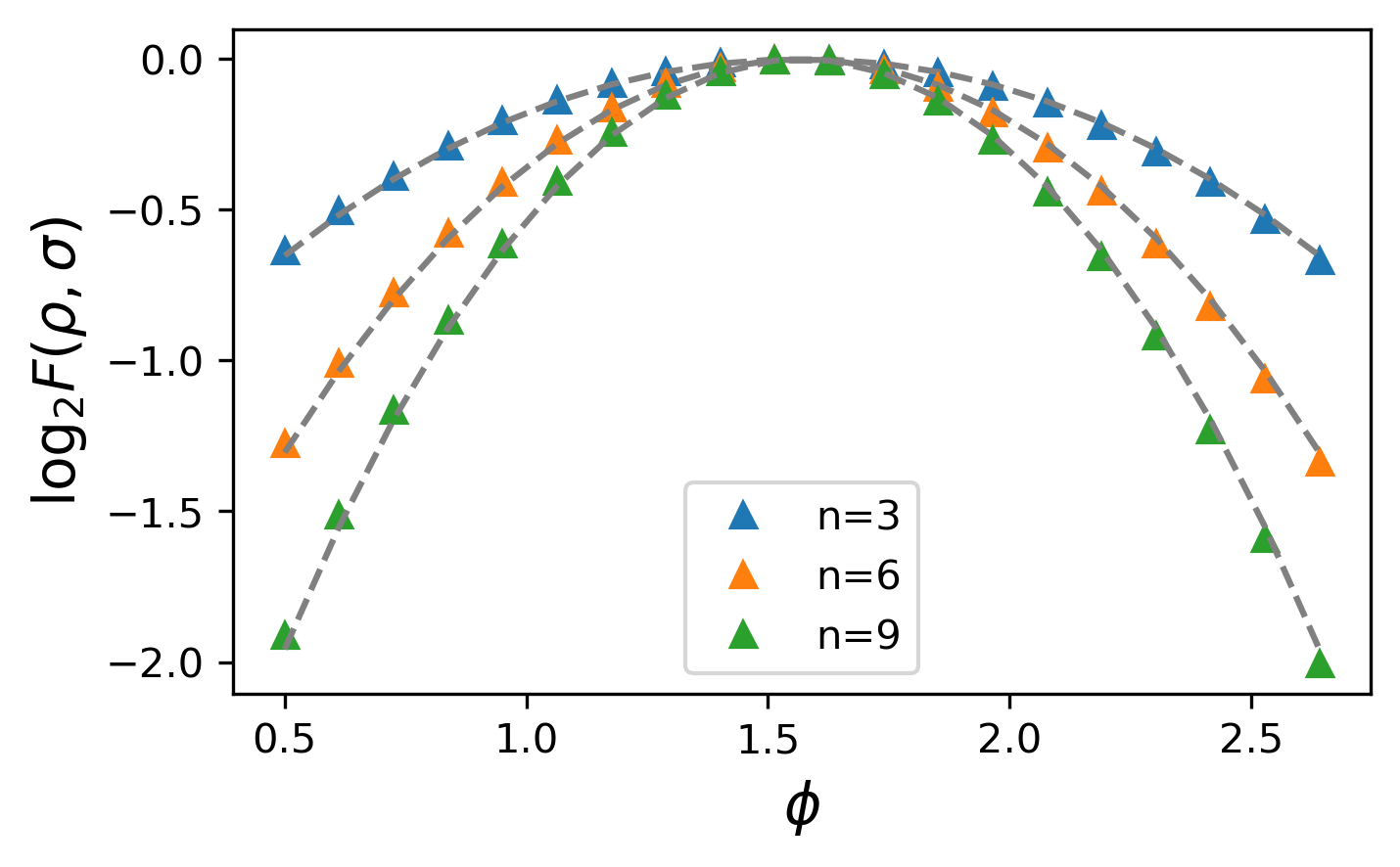}

	\end{subfigure}
	\caption{ Exact (dashed lines) and estimated (triangle points) values of the (left) rank, (middle) R{\'e}nyi entropy for $\alpha = 1/2$, and (right) fidelity for $n=3, 6, 9 $ qubit states. The input state is a product state given by $\rho = (\cos^2(\phi/2) \ketbra{0} +\sin^2(\phi/2) \ketbra{1})^{\otimes n}$. For the fidelity calculation $F(\rho,\sigma)$, the second input state was chosen to be the maximally mixed state $\sigma = \openone_d/d$, where $d=2^n$. Here, the cost function is directly computed rather than obtained via simulated quantum circuits as in Fig.~\ref{fig::rankEntropy}.}
	\label{fig::rankEntropyMultiQubit}
\end{figure*}

A limitation of simulating actual quantum circuits is that it is very memory intensive, which limits our ability to scale up to larger system sizes. In order to scale the proposed algorithms, we numerically evaluate the cost function Eq.~(\ref{eq::costFun}) directly instead of simulating actual quantum circuits. The gradient descent is otherwise performed just as before. The input states are product states $\rho = (\cos^2(\phi/2) \ketbra{0} +\sin^2(\phi/2) \ketbra{1})^{\otimes n }$ where $n=3,6,9$ are the number of qubits. The ansatz state is chosen to be $\eta(\theta) = (\cos^2(\theta/2) \ketbra{0} +\sin^2(\theta/2) \ketbra{1})^{\otimes n}$, initialized at $\theta = \pi/2$. The simulated results are shown in Fig.~\ref{fig::rankEntropyMultiQubit}, where it is observed that the proposed VQAs give highly accurate estimates of the rank, R{\'e}nyi entropy, and fidelity. Over the range of parameters tested, the errors do not appear to grow exponentially with system size, which is a major problem faced by many VQAs. For example, for $n=3,6,9$, the mean errors of the rank estimation are $0.0348\%, 0.0695\%$ and $0.104\%$ respectively, which approximately follows a linear scaling. However, because the rank of $\rho$ grows exponentially in this example, we expect errors to be a much more significant issue for larger system sizes. In general, the scaling of errors with system size is highly dependent on both the rank of the input state as well as the design of the ansatze. This will be discussed in greater detail in the following section.

\section{Strategies for mitigating the barren plateau landscape problem}

The cost function in Eq.~(\ref{eq::costFun}) in involves measurements of global properties, which may lead to the Barren Plateau Landscape (BPL) problem where the expected magnitude of the gradient of the cost function decreases exponentially with system size \cite{mcclean2018barren,cvbpl}.

For example, suppose the input state is a tensor product state of the form $\rho = \begin{pmatrix}
\lambda_1 & 0 \\
0 & \lambda_2
\end{pmatrix}^{\otimes n}$, $\lambda_{1}>\lambda_{2}$ and that the ansatz state has the form $\eta(\theta) = R(\theta) \rho^{-1/(2k)}/\tr(\rho^{-1/(2k)}) R(\theta)^\dag $, where $\theta = (\theta_1,\ldots, \theta_n)$ and $R(\theta) = \bigotimes_{i=1}^n \exp(-i\theta_i \sigma_y/2)$. Despite the fact that $\eta(\theta)$ is an SU(2) orbit of the state that achieves the global minimum of Eq.~(\ref{eq::costFun}), it can be shown that this combination of input state $\rho$ and ansatz $\eta(\theta)$ encounters the BPL problem. To see this, first note that the cost function can be written in the form \begin{equation}
    C(\theta)={1\over 2^{n}}\prod_{j=1}^{n}{N(\theta_{j})\over D(\theta_{j})}
    \label{eqn:shortcf}
\end{equation}
where $N(\theta_{j})$ and $D(\theta_{j})$ are quartic polynomials in $\tan{\theta_{j}\over 2}$ with coefficients that depend on $\lambda_{1},\lambda_{2}$. From independence of the random variables $\theta_{j}$ distributed according to the uniform distribution $p$ on $[-\pi,\pi]$, we note that 
\begin{align}
\underset{\theta\sim p^{\times n}}{E}\left( \vert \del_{\theta_{1}}C(\theta) \vert \right) &= 2^{-n}\underset{\theta_{1}\sim p}{E}\left( \vert\del_{\theta_{1}}{N(\theta_{1})\over D(\theta_{1})} \vert\right) \left( \underset{x\sim p}{E}\left(  {N(x)\over D(x)} \right) \right)^{n-1}
\label{eqn:rrr}
\end{align}
where we considered optimization with respect to $\theta_{1}$ without loss of generality.
The first factor of Eq.~(\ref{eqn:rrr}) is independent of $n$, and when $p$ is restricted to the neighborhood $[-\delta,\delta]$ for $\delta \ll {\lambda_{2}\over \lambda_{1}}$, the second factor is given by \begin{equation}\left( \left({1\over 2}-{\delta^{2}\left( {\lambda_{2}\over \lambda_{1}}+{\lambda_{1}\over \lambda_{2}}  \right) \over 48} \right) + O(\delta^{2}) \right)^{n}, \label{eqn:bobo}\end{equation} which has an exponential dependence on $n$. Therefore, from Chebyshev's inequality in the form $P(\vert X\vert >\epsilon)\le \epsilon^{-1}{E(\vert X\vert)}$, it follows that on $[-\delta,\delta]$, $P(\vert \del_{\theta_{1}}C(\theta) \vert \ge \epsilon ) \le O(b^{n})$ where $b$ is taken from Eq.~(\ref{eqn:bobo}). This demonstrates that in some local neighborhood of the optimal solution, the gradient exponentially decays with the system size $n$. The exponential decay of the gradient magnitude with respect to $n$ is the defining property of BPL.

\begin{figure*}[t]
\captionsetup[subfigure]{justification=centering}
    \centering
	\begin{subfigure}[c]{0.4\textwidth}
\includegraphics[scale=0.4]{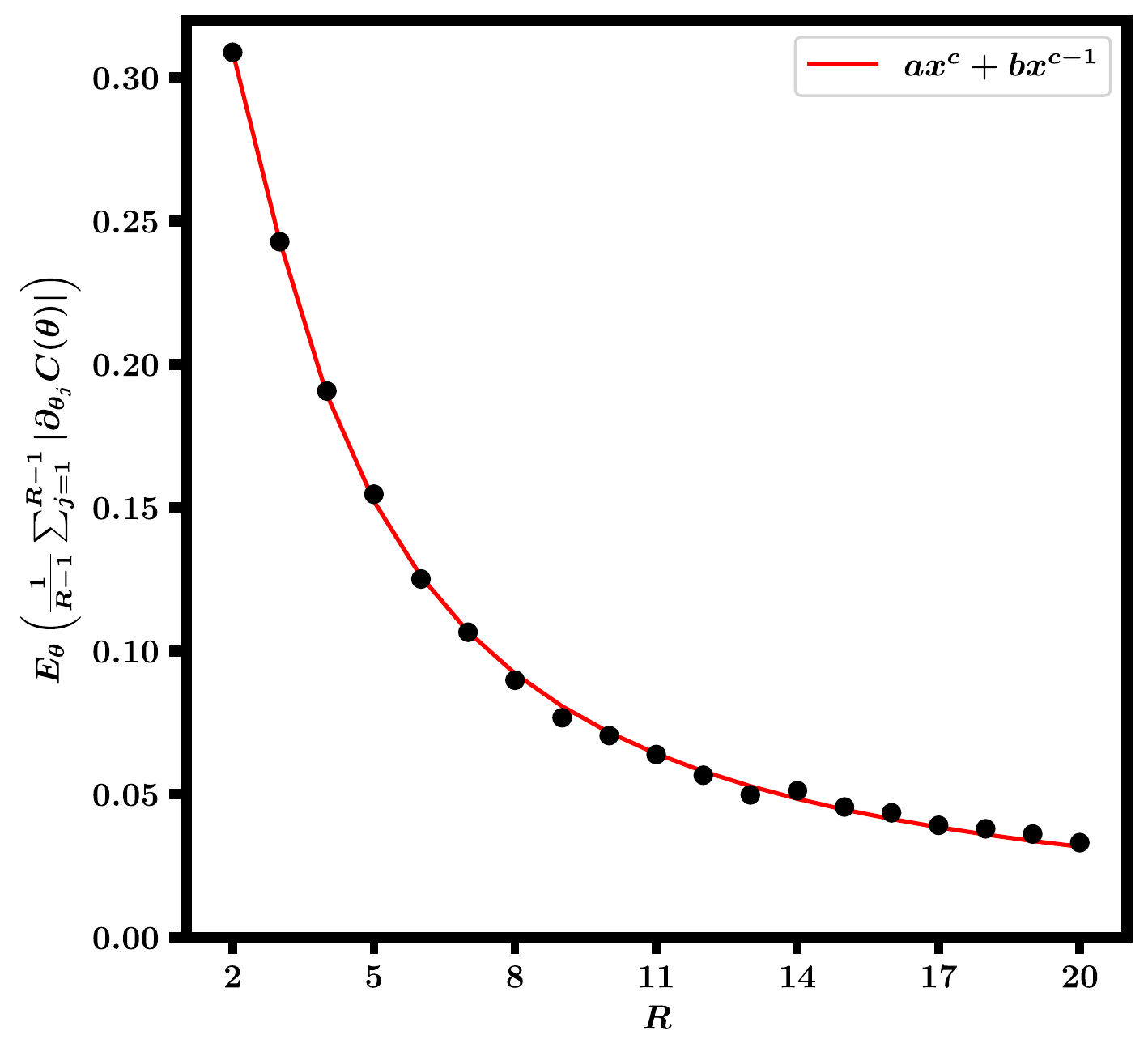}
\end{subfigure}
\begin{subfigure}[c]{0.4\textwidth}
\includegraphics[scale=0.4]{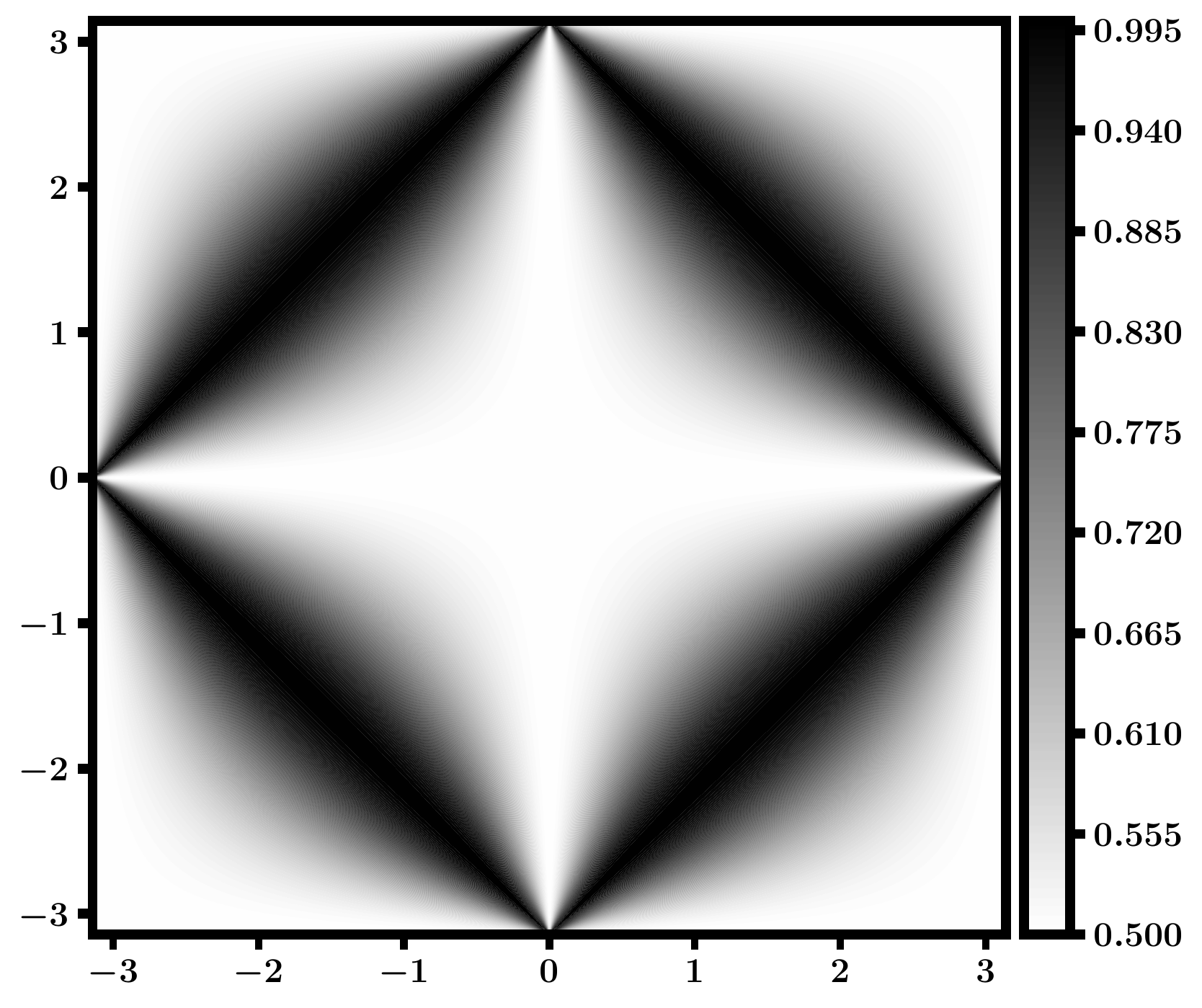}
\end{subfigure}
	\caption{(Left) Average magnitude of the gradient components of cost function (\ref{eq::costFun}) for $k=4$ and for $R=2,3,\ldots, 20$ and $\rho$ the completely mixed state of rank $R$. Error bars are smaller than the dot radius. The red curve is a least squares fit of the function $f(x)=ax^{c}+bx^{c-1}$ to the data; optimal parameters $a\approx 1.41$, $b\approx -1.35$, $c\approx -1.25$.  (Right) Grayscale plot of $C(\theta)$ defined by (\ref{eqn:nnnn}) and (\ref{eqn:dddd}) with $\lambda=1/2$, $\mu=\lambda^{1/2k}/(\lambda^{1/2k}+(1-\lambda)^{1/2k})$, and $n=2$.\label{fig:ttt}. The gradient vanishes with $n$ in the center region near the solution. }
\end{figure*}

One possible way to mitigate the BPL problem is the introduction of additional local terms to the cost function \cite{cfdbp}. We demonstrate this possibility for at least some choices of the state $\rho$. 
Let us now denote the cost function from Eq.~(\ref{eq::costFun}) as $C_G(\theta)$ to reflect the global property of the cost function. We consider additional local terms of the form:
\begin{align}
C_L(\theta) &= \frac{1}{n}\sum_i  \tr ( \sigma_i^2),
\label{eqn:si}
\end{align} where $n$ is the number of subsystems of $\rho$ and $\sigma_i = \eta_i(\theta)^{k} \rho_i \eta_i(\theta)^{k}/\tr(\eta_i(\theta)^{k} \rho_i \eta_i(\theta)^k)$. The states $\eta_i(\theta) = R_i(\theta_i) \rho_i^{-1/(2k)}/\tr(\rho_i^{-1/(2k)}) R_i(\theta_i)^\dag$ and $\rho_i = \begin{pmatrix} \lambda_1 & 0 \\ 0 & \lambda_2 \end{pmatrix}$ are taken to be the reduced density matrices of the $i$th subsystem of $\eta(\theta)$ and $\rho $, respectively. Observe that unlike $C_G(\theta)$, each term in $C_L(\theta)$ measures a local property of the state and scales at most polynomially with system size.

We then observe that in the example we have considered, $\tr ( \sigma_i^2)$, and hence $C_L(\theta)$, is minimized when $\theta_i = 0$. This is the same critical point as the optimal solution for the global cost function $C_G(\theta)$. This allows us to circumvent the BPL problem. More generally however, it is not always true that the solution to the local and the global cost functions coincide. As a possible BPL mitigation strategy, one approach is to consider a linear combination of both local and global cost functions of the form $C_{G+L}(\theta) = (1-\epsilon)C_G(\theta)+\epsilon C_L(\theta)$, where $\epsilon \ll 1$ is a small constant. By choosing small $\epsilon$, we ensure that we are close to the ideal solution in the worst case, while eliminating the BPL problem in cases such as the previously considered examples. To see that we will always be close to the actual solution for small $\epsilon$, we observe that $\lVert \openone_{d}/d - \sigma \rVert^2 < C_G(\theta) <C_{G+L}(\theta) \lesssim \epsilon $ (see Eq.~(\ref{eq::geomInt})). In the last inequality, we assumed the approximation $C_G(\theta) \approx 0$ when the dimension $d$ is sufficiently large and the inequality $C_L(\theta) \leq 1$.

Another possible strategy to mitigate the BPL problem is to consider correlated ansatz parameters. Because $\rho$ is an exchangeable state \cite{watrousbook}, one expects that efficient optimization of the cost function $C_{G}(\theta)$ can be achieved by correlation of the ansatz parameters in $R(\theta)$, e.g., by taking $\theta_{j}=\theta$ for all $j$  \cite{vc}. However, we observe that the $2^{-n}$ factor in Eq.~(\ref{eqn:shortcf}) cannot be overcome by straightforwardly correlating the parameters $\theta_{j}$, so an alternative strategy is required. In Appendix \ref{app:aa}, we show that BPL still exists for this scenario. In the case that $\rho$ is an unnormalized state in Eq.(\ref{eq::costFun}), e.g., $\rho=\sigma^{\ell}$ for quantum state $\sigma$, the purity minimization results in the state proportional to $\sigma^{-\ell / 2k}$. Then the cost function for a correlated $R(\theta)$, $\sigma=\text{diag}(\lambda_{1},\lambda_{2})$, and $\eta(\theta)\propto R(\theta)\sigma^{-\ell/2k}R(\theta)^{\dagger}$ is a function of $\theta$ and $\ell$. For any $\ell$, one can demonstrate BPL in a neighborhood of the cost function minimum by using the same method that is used to show the existence of local BPL for $C_{G}(\theta)$ in Appendix \ref{app:aa}.

We now provide evidence that for low rank states, the BPL problem can be avoided with sufficiently good ansatz. We show that if an $n$-qubit state $\rho$ has rank $R$, then there exists an ansatz $\eta(\theta)$ defined by a parameter manifold of dimension $R-1$ (specifically, the $R$-sphere) for which BPL does not occur for the cost function Eq.~(\ref{eq::costFun}) if $R=\text{poly}(n)$. Consider the ansatz \begin{align}\eta(\theta)&=\cos^{2}{\theta_{1}\over 2}\ket{\psi_{1}}\bra{\psi_{1}} +\sin^{2}{\theta_{1}\over 2}\cos^{2}{\theta_{2}\over 2}\ket{\psi_{2}}\bra{\psi_{2}}\nonumber \\
&+ \ldots + \prod_{j=1}^{R-1}\sin^{2}{\theta_{j}\over 2} \ket{\psi_{R-1}}\bra{\psi_{R-1}} ,\label{eqn:suffans}\end{align} where $\lbrace\ket{\psi_{i}}\rbrace_{i=1}^{R}$ are the eigenvectors of $\rho$. To show that the resulting cost function Eq.~(\ref{eq::costFun}) does not exhibit BPL if $R=\text{poly}(n)$, it is sufficient to show that the expected average magnitude of the gradient components $E\left( {1\over R-1}\sum_{j=1}^{R-1}\vert \del_{\theta_{j}}C(\theta)\vert \right)$ goes to 0 polynomially in $R$. We demonstrate this numerically in Fig.\ref{fig:ttt} by fitting the mean of  Monte Carlo estimates of the expected gradient components to a polynomially decreasing function. 

More generally, low rank states combined with low rank variational ansatze $\eta(\theta)$ do not guarantee the absence of BPL in a given problem. 
Consider cost function (\ref{eq::costFun}) with rank 2 state $\rho=\lambda \ket{0}\bra{0}^{\otimes n}+(1-\lambda)\ket{1}\bra{1}^{\otimes n}$, and with ansatz $\eta(\theta):= R(\theta) \sigma  R(\theta)^{\dagger}$, where the normalized state $\sigma$ is defined by
\begin{equation}
    \sigma= (1-\mu)\ket{0}\bra{0}^{\otimes n} + \mu\ket{1}\bra{1}^{\otimes n},
    \label{eqn:sigsig}
\end{equation}
$\mu \in (0,1)$. For simplicity, we consider fixed $\lambda$ and $\mu$, viz., the spectra of the target $\rho^{-1/2k}$ and the ansatz $\eta(\theta)$ do not vary with $n$.
Unlike the ansatz (\ref{eqn:suffans}), $\eta(\theta)=R(\theta)\sigma R(\theta)^{\dagger}$ does not commute with $\rho$ pointwise over $\theta$. Considering $C(\theta)$ as a function of $\lambda$, $\mu$ and $\theta$, one has that $C(\theta)\ge {1\over 2}$, as expected for a state $\rho$ in a 2 dimensional subspace. The cost function is given by $C(\theta)=N(\theta)/D(\theta)$ with
\begin{align}
    N(\theta)&= \lambda^{2}\left((1-\mu)^{2k}a(\theta) + \mu^{2k} b(\theta)\right) \nonumber \\
    &{} + 2\lambda(1-\lambda)((1-\mu)^{2k}+\mu^{2k})^{2}a(\theta)b(\theta) \nonumber \\
    &{} + (1-\lambda)^{2}\left( (1-\mu)^{2k}b(\theta) + \mu^{2k} a(\theta)\right)
    \label{eqn:nnnn}
\end{align}
and
\begin{align}
    D(\theta)&=\left( \lambda\left((1-\mu)^{2k}a(\theta) + \mu^{2k} b(\theta)\right) \right. \nonumber \\ 
    &+ \left. (1-\lambda)\left( (1-\mu)^{2k}b(\theta) + \mu^{2k} a(\theta)\right) \right)^{2},
    \label{eqn:dddd}
\end{align}
where $a(\theta):= \prod_{j=1}^{n}\cos^{2}{\theta_{j}\over 2}$, $b(\theta):=\prod_{j=1}^{n}\sin^{2}{\theta_{j}\over 2} $.
The cost function landscape is shown Fig. \ref{fig:ttt}, with $\lambda=1/2$ and $\mu=\lambda^{1/2k}/(\lambda^{1/2k}+(1-\lambda)^{1/2k})$ so that $C(\theta=0)=1/2$, i.e., the ansatz is fully expressive. One finds that $\vert \del_{\theta_{1}}C(\theta)\vert =  F(\lambda,\mu) {\vert\theta_{1}\vert\over 4^{n-1}} \prod_{j=2}^{n}\theta_{j}^{2} +O(\delta^{4n})$ with constant $F(\lambda,\mu)$ in a sufficiently small neighborhood $(-\delta,\delta)^{\times n}$, $\delta < 1$, of $\theta=0$. It follows that BPL is present in a neighborhood of a global minimum. In fact, for the fully expressive ansatz in Fig.\ref{fig:ttt} we find that BPL exists on sufficiently small constant volume neighborhoods, e.g., $(-2,2)^{\times n}$. However, similarly to the above analysis of a correlated ansatz for a full rank state, for a given $\lambda$, there are $\mu$ for which BPL is absent on the global domain $[-\pi,\pi]^{\times n}$. The presence of local BPL  even though there is no global BPL present suggests that this low rank ansatz was only able to partially mitigate the BPL problem. Therefore, although it is possible that a randomly initialized gradient descent algorithm could rapidly obtain a cost function value close to the global minimum, the ansatz $\eta(\theta)$ will require exponential resources to completely converge to the solution $\rho^{-1/2k}/\text{tr}\rho^{-1/2k}$ in terms of trace distance, even if the ansatz is fully expressive.

\section{Conclusion}

We proposed VQAs to estimate physical quantities such as the rank, R{\'e}nyi and Tsallis quantum entropies, the quantum fidelity and the quantum Fisher information for mixed quantum states. These VQAs share a common theme of minimizing a cost function related to the quantum purity of a normalized quantum state. In general, these cost functions are efficiently computable on quantum computers via a quantum SWAP test, while it may be inefficient on a classical computer. We also apply similar cost functions to perform tasks such as preparing fractional powers and fractional inverses of quantum states, as well as quantum state learning. It is noteworthy that already, a rather wide variety of problems can be solved via the purity minimization approach, which may suggest further applications beyond those we have considered thus far. As a proof of concept, numerical simulations of the VQAs were performed which demonstrates that the proposed algorithms are able to retrieve the exact values for the rank, R{\'e}nyi and quantum fidelity for up to $n=9$ qubit states.

We also studied the BPL problem in relation to the basic cost function that we proposed in Eq.~(\ref{eq::costFun}). Explicit examples exhibiting the BPL problem  were discussed, and strategies such as adding local terms to the cost function or correlating the ansatz parameters were considered in order to mitigate this issue. Our numerical evidence further suggests that for low rank input states, there exist low dimension ansatz manifolds for which BPL does not occur. This supports the recent results in Ref.\cite{Wang2021} which suggests that for high rank states, computing the quantum fidelity may be hard even for quantum computers.

We are hopeful that the methods proposed here will inspire new applications for VQAs and for quantum computing in general. In particular, we are optimistic that some of the proposed VQAs can be applied on already available NISQ computing devices to probe nonclassical properties of quantum states. 

{Acknowledgments.---} K.C.T. was supported by the NTU Presidential Postdoctoral Fellowship program funded by Nanyang Technological University. T.V. acknowledges support from the LDRD program at LANL.

\bibliography{refs.bib}

\appendix
\section{BPL for ansatz $\eta(\theta)$ with correlated parameters\label{app:aa}}

Correlating the angles in $R(\theta_{1},\ldots,\theta_{n})$ via $\theta_{j}=\theta$ for all $j$ gives the variational ansatz
\begin{equation}
    \eta(\theta)= \left( e^{-i{\theta\over 2}\sigma_{y}} \right)^{\otimes n}{\rho^{-1/(2k)}\over \tr(\rho^{-1/(2k)})} \left( e^{i{\theta\over 2}\sigma_{y}} \right)^{\otimes n}.
\end{equation}The cost function (\ref{eqn:shortcf}) then takes the form 
\begin{align}
C(\theta)&={1\over 2^{n}}\left( N(\theta)\over D(\theta) \right)^{n}\nonumber \\
N(\theta)&= 1+b\tan^{2}{\theta\over 2}+a\tan^{4}{\theta\over 2} \nonumber \\
D(\theta)&= \left(1+c\tan^{2}{\theta\over 2}\right)^{2}
\end{align}
 where $a:= {1\over 2}({\lambda_{1}^{2}\over \lambda_{2}^{2}}+{\lambda_{2}^{2}\over \lambda_{1}^{2}})$, $b:=2\left( {\lambda_{2}\over \lambda_{1}}+{\lambda_{1}\over \lambda_{2}}  -1\right)$, $c:={1\over 2}\left( {\lambda_{2}\over \lambda_{1}}+{\lambda_{1}\over \lambda_{2}}  \right)$. We seek the $n\rightarrow \infty$ asymptotics of
$E_{\theta \sim p}\left(  \vert \del_{\theta}C(\theta) \vert \right)$,
in particular an upper bound that vanishes exponentially. We restrict the uniform measure $p$ to the uniform measure on $A_{\delta}=[-\delta,\delta]$ where $\delta \ll \sqrt{\lambda_{2}\over\lambda_{1}}$ and prove that  BPL occurs in this  neighborhood of the optimum.
\begin{align}
E_{\theta \sim p}\left(  \vert \del_{\theta}C(\theta) \vert \right)&= {n\over 2^{n}}E_{\theta \sim p}\left(  \left( N(\theta)\over D(\theta) \right)^{n-1}\Big\vert \del_{\theta}{N(\theta)\over D(\theta)} \Big\vert \right)\label{eqn:corrcorr}
\end{align}
It follows that
\begin{align}
\Big\vert \del_{\theta}{N(\theta)\over D(\theta)} \Big\vert &= \Big\vert { ((2a-bc)\tan^{2}{\theta\over 2} + b-2c)\tan{\theta\over 2} \over (1+c\tan^{2}{\theta\over 2})^{3}\cos^{2}{\theta\over 2} } \Big\vert \nonumber \\
&\sim  \Big\vert { (b-2c){\theta\over 2} \over (1+c{\theta^{2}\over 4})^{3}}\Big\vert +O(\theta^{3}) \; \; (\text{ for }\theta \rightarrow 0)
\end{align}
So we calculate 
\begin{align}
E_{\theta \sim p}\left(  \vert \del_{\theta}C(\theta) \vert \right) &= {\vert b-2c\vert n\over 2^{n+1}}{1\over 2\delta}\int_{-\delta}^{\delta}dx {\vert x\vert \over (1+{cx^{2}\over 4})^{n+2}}+O(\delta^{3})\nonumber \\
&= {\vert b-2c\vert n\over c(n+1)2^{n}}\left(1-(1+{c\delta^{2}\over 4})^{-(n+1)} \right).
\end{align}
Again using the Chebyshev inequality, the probability of an arbitrarily small gradient magnitude scales as $O(2^{-n})$, so there is barren plateau even when an ansatz with correlated parameters is used.



\end{document}